\documentclass[runningheads]{llncs}
\usepackage[utf8]{inputenc}
\usepackage{lmodern}
\usepackage[T1]{fontenc}
\usepackage{graphicx}
\usepackage{csquotes}

\usepackage{booktabs}
\usepackage{tabularx}
\usepackage{multirow}

\usepackage{tcolorbox}
\tcbuselibrary{theorems}
\usepackage{fontawesome}

\newtcbtheorem{user}{\faWechat{} User Prompt}
{colback=blue!5,colframe=blue!35!black,fonttitle=\bfseries}{th}

\newtcbtheorem{assistant}{\faWechat{} Assistant}
{colback=blue!20,colframe=blue!35!black,fonttitle=\bfseries}{th}

\begin{document}
\title{Using Large Language Models for Natural Language Processing Tasks in Requirements Engineering: A Systematic Guideline}
\titlerunning{LLMs for NLP Tasks in RE: A Systematic Guide}
\author{
Andreas Vogelsang\inst{1}\orcidID{0000-0003-1041-0815}
\and
Jannik Fischbach\inst{2,3}\orcidID{0000-0002-4361-6118} 
}
\authorrunning{A. Vogelsang and J. Fischbach}
\institute{
University of Cologne, Cologne, Germany\\
\email{vogelsang@cs.uni-koeln.de}
\and
Netlight Consulting GmbH, Munich, Germany \and 
fortiss GmbH, Munich, Germany\\ 
\email{jannik.fischbach@netlight.com}
}
\maketitle              % typeset the header of the contribution
\begin{abstract}
% Context and Motivation
Large Language Models (LLMs) are the cornerstone in automating Requirements Engineering (RE) tasks, underpinning recent advancements in the field. Their pre-trained comprehension of natural language is pivotal for effectively tailoring them to specific RE tasks. However, selecting an appropriate LLM from a myriad of existing architectures and fine-tuning it to address the intricacies of a given task poses a significant challenge for researchers and practitioners in the RE domain. Utilizing LLMs effectively for NLP problems in RE necessitates a dual understanding: firstly, of the inner workings of LLMs, and secondly, of a systematic approach to selecting and adapting LLMs for NLP4RE tasks.
% Content
This chapter aims to furnish readers with essential knowledge about LLMs in its initial segment. Subsequently, it provides a comprehensive guideline tailored for students, researchers, and practitioners on harnessing LLMs to address their specific objectives.
% Contribution
By offering insights into the workings of LLMs and furnishing a practical guide, this chapter contributes towards improving future research and applications leveraging LLMs for solving RE challenges.
\end{abstract}
\section{Introduction}

Large Language Models (LLMs) have revolutionized how we can automate Requirements Engineering (RE) tasks and what quality we can achieve. Besides the improved quality over traditional classification algorithms, LLMs are mainly successful in RE because they do not need extensive datasets to be trained. RE researchers and practitioners have been limited for a long time by the scarcity of high-quality data for specific RE tasks. Early studies on deep learning for RE tasks have shown that simple tasks such as binary requirements classification could only be solved effectively with labeled datasets of sizes larger than 10,000 entries~\cite{Winkler2016}. With the rise of LLMs and especially the distribution of pre-trained models, RE researchers can adapt the models to certain domains and tasks with only a few additional and specific data~\cite{Hey2020}.

LLMs belong to an active research field with many more active researchers than the field of Requirements Engineering. RE researchers and practitioners can profit from the recent advances in models and their accessibility. Examples of RE tasks that have been successfully (semi-)automated include requirements classification~\cite{Hey2020}, requirements tracing~\cite{lin2021traceability}, test case derivation~\cite{Fischbach2023}, or requirements completion~\cite{luitel2023improving}.
On the other hand, RE researchers need to stay on top of the development of modern NLP techniques and be able to decide and assess which improvements may help them solve RE tasks. 

Adopting LLMs in RE marks a significant milestone in the quest for automation and quality enhancement within the domain. However, amid the enthusiasm surrounding their efficacy, the selection of suitable LLM architectures and the nuanced process of fine-tuning them to meet the specific demands of RE tasks remains challenging. This challenge is multifaceted, rooted in the complexity of LLM architectures and compounded by the intricate nature of RE requirements.

LLMs encompass various architectures, each tailored to address different facets of natural language understanding and generation. From encoder-only models like BERT to generative models like GPT, the landscape of LLMs is vast and continually evolving. Consequently, RE researchers and practitioners are confronted with the daunting task of discerning which architecture aligns best with the intricacies of their respective tasks.

Moreover, fine-tuning LLMs to optimize performance for RE tasks requires a delicate balance between general language comprehension and domain-specific adaptation. Unlike generic NLP tasks, RE demands a nuanced understanding of domain-specific jargon, context, and semantic nuances. Thus, the fine-tuning process necessitates meticulous calibration to preserve the fidelity of the model's pre-trained knowledge while tailoring it to the idiosyncrasies of RE requirements.

This chapter aims to provide RE researchers and practitioners with a fundamental understanding of what LLMs are and how they work (see Section~\ref{sec:fundamentals}). Moreover, we show three fundamentally different ways of working with LLMs to solve RE tasks (see Section~\ref{sec:guideline}). We close the chapter with a summary and concluding remarks (see Section~\ref{sec:conclusion}).

\section{Fundamentals of LLMs}
\label{sec:fundamentals}

\subsection{Sparse, Static, and Contextual Word Embeddings}
\label{sec:embeddings}
This section describes the idea of word embeddings and the difference between static and contextual word embeddings that build the foundation of LLMs.

\textbf{Input Representation.} To process Natural Language (NL) using computational models, NL data needs to be transformed into a vectorized representation~\cite{Bengio13}. Pursuing effective representations for NL data has engaged researchers for several decades~\cite{blitzer06,Mikolov13,penningto14}. The concept of employing a vector space to encapsulate the semantic essence of words traces back to Osgood et al.\cite{Osgood57}. A mapping of a word to its corresponding vector is also called \textit{word embedding} (i.e., a word is encoded in the vector space). The primary objective of word embeddings is to encapsulate a word's meaning based on its contextual usage. In particular, \enquote{words appearing in \textit{similar contexts} tend to possess \textit{similar meanings}}~\cite{Jurafsky2009} and should therefore be mapped to similar locations in vector space. Consequently, embeddings strive to characterize words through their contextual companions.

\textbf{Sparse Embeddings.} Based on the idea that the similar distribution of words is associated with similar meaning (\textit{distributional hypothesis}~\cite{Firth57}), various methods have been developed for generating word embeddings. For instance, the \textit{Term Frequency–Inverse Document Frequency} approach relies on straightforward co-occurrence statistics to construct word embeddings by counting nearby words. Consequently, the vectors' dimensionality corresponds to the vocabulary size employed. This results in word embeddings that are excessively long and dominated by zeros, as many words rarely appear near others~\cite{Jurafsky2009}. This type of word embedding is often described as \textit{sparse}.

\textbf{Static Embeddings.} Multiple techniques have been devised to create \textit{dense} word embeddings to mitigate the problem of embedding size ballooning as the vocabulary expands. These techniques include Word2Vec~\cite{Mikolov13}, Global Vectors for Word Representation (GloVe)~\cite{penningto14}, and FastText~\cite{bojanowski16}. In contrast to sparse embeddings, dense vectors have fixed dimensions and consist of real-valued numbers, which can be either positive or negative. Word2Vec, for instance, employs a classifier to predict the likelihood of two words being neighbors and then utilizes the weights of the trained classifier as word embeddings. The resulting embeddings are \textit{static}, meaning that the model generates a single unchanging vector for each word in the vocabulary. Nevertheless, static embeddings have a clear limitation: a word is invariably represented by the same vector, irrespective of the context in which it appears. This constraint restricts their capacity to capture the multifaceted meanings of words, as words naturally adopt different meanings based on their context.

\textbf{Contexual Embeddings.} LLMs address this shortcoming by considering word context when generating embeddings. In essence, contemporary language models estimate the likelihood of a given word sequence occurring and employ the computed hidden states as word embeddings. Consequently, the contextualized embedding of a word is contingent on all other words within a sentence (see self-attention mechanism in Section~\ref{sec:trans-architecture}). This enables the creation of distinct word embeddings for the same word. To generate contextualized word embeddings, it is imperative to transform the input sequence into a suitable format. To this end, the sequence undergoes segmentation into individual components, a process facilitated by tokenizers such as the \textit{WordPiece}~\cite{wordpiece} tokenizer. The \textit{WordPiece} tokenizer operates on a subword level, intending to encompass an extensive vocabulary using a finite set of established words. Specifically, \textit{WordPiece} aims to overcome the limitations of tokenization methods reliant on whole words, which often struggle to handle out-of-vocabulary (OOV) words. \textit{WordPiece} adopts a strategy where frequently used words remain unaltered as whole tokens, while less common words are broken down into subword units recognized by the model. 

Consequently, even OOV words can be processed, as their constituent subwords often retain sufficient semantic information to enable the model to deduce the intended meaning of the OOV word. For instance, consider the application of the \textit{WordPiece} tokenizer to the term associated with the foundational architecture of modern LLMs: ``Transformer''. In this case, the word ``Transformer'' is classified as infrequent and is thus segmented into ``Transform'' and ``\#\#er'', where the ``\#\#'' prefix indicates that this token is an extension of the preceding one. Despite being unfamiliar to the model, the semantics of ``Transformer'' are intelligible due to the presence of a commonly used word, the verb ``transform''.

\subsection{Transformer Architecture}
\label{sec:trans-architecture}
This section describes the inner workings of the transformer architecture on which modern LLMs are based. In its original configuration, a transformer architecture comprises two primary components: an encoder, which converts an input (e.g., German text) into a vectorized representation, and a decoder, responsible for transforming this vectorized representation into an output (e.g., English text). This structural division is rooted in the transformer's initial purpose of facilitating machine translation~\cite{NIPS2017}. To understand the inner workings of the transformer architecture, we must study the main components embedded in each encoder and decoder: self-attention mechanism, multi-head attention, positional encoding, and layer normalization. A great visualization of the interaction of all components can be found on Jay Alammer's blog\footnote{\url{https://jalammar.github.io/illustrated-transformer/}}.

\textbf{Positional encoding.} A transformer adds positional encoding to the input embeddings to provide information about the position of each token in the sequence. This allows the model to understand the sequential order of the input data, which is crucial for tasks such as language understanding.

\textbf{Self-attention mechanism.} Self-attention enables the model to weigh the importance of different words in a sequence relative to each other. This mechanism allows the model to consider the entire context rather than processing words sequentially. Specifically, the self-attention mechanism computes attention scores for each word based on its relationships with other words, creating a weighted sum representing each word's context.

\textbf{Multi-head attention.} Multi-head attention runs the self-attention mechanism in parallel multiple times (heads) and concatenates the results. Each head attends to different aspects of the input, enabling the model to learn richer features. The concatenated outputs are linearly transformed to produce the final multi-head attention output.

\textbf{Feedforward neural network.} After the multi-head attention mechanism, a feedforward neural network processes the information learned through the attention mechanism. This network consists of fully connected layers and activation functions, introducing non-linearity to the model and enabling it to capture complex relationships within the data.

\textbf{Layer normalization.} Layer normalization is applied after each sub-layer, such as the self-attention or feedforward layers, to stabilize the learning process. It normalizes the values within a layer, preventing the model from amplifying undesired features or gradients during training. This contributes to the stability and efficiency of the transformer architecture.

\subsection{Encoder-only LLMs}

BERT is an \textit{encoder-only} model and focuses on the derivation of contextual embeddings. Specifically, it maps an input sequence $x_1 \dots x_n$ to a contextualized encoded sequence $y_1 \dots y_n$:

\begin{equation*}
    f_{\text{BERT}} : x_1 \dots x_n \mapsto y_1 \dots y_n
\end{equation*}

BERT consists of a stack of multiple encoders. It is available in two versions: the \textbf{BERT-base} model has 12 encoder layers stacked on each other, whereas the \textbf{BERT-large} model includes 24 encoder layers. The distinctive feature of BERT compared to other language models is its \textit{bidirectionality}. Specifically, it considers a word's left and right context when creating a corresponding embedding~\cite{devlin19}. BERT is trained by optimizing two tasks: Masked language modeling (MLM), where a portion of the input tokens in a sentence are masked and the model needs to predict the masked tokens, and Next Sentence Prediction (NSP), where the model learns to predict whether two sentences follow each other. Previous transformer models like GPT-2 process an input sequence only in a left-to-right fashion. Hence, the hidden states are computed independently of the others as the model only considers tokens seen earlier in the context (i.e., the embeddings only contain information of the \textit{right} context). This issue is especially problematic when utilizing embeddings to solve complex NLP problems (e.g., fine-grained sentiment analysis and question answering). Using the self-attention mechanism enables BERT to understand long-range dependencies between tokens and, thus, to overcome the vanishing gradient problem of sequential models. To this end, BERT applies the self-attention mechanism over the entire input sequence and contextualizes each token using information from the entire input. The final embeddings have a size of 768 dimensions (\textbf{BERT-base}) or 1024 (\textbf{BERT-large}). The dense layer within the encoder is used to further enrich the attention mechanism's output and pass it to the next encoder.

\subsection{Decoder-only (Generative) LLMs and Prompting}
\label{sec:prompting}

Decoder-only LLMs have been designed to generate text. To support the generative capabilities of decoder-only LLMs, they are primarily pre-trained with a next-word prediction (NWP) objective, where the models predict the next word or words in a given sequence of words. After pre-training, decoder-only LLMs are triggered by a so-called \textit{prompt}. A prompt is a textual input instructing the generative LLM to generate the desired response.
Feeding decoder-only LLMs with prompts offers a new paradigm for interaction. In contrast to non-generative LLMs, it is not necessary to encode information about the task and the input in a smart way. Instead, the task and the input can be expressed in natural language and passed directly to the model. The model's output is richer than non-generative LLMs since it produces (customizable) text instead of confidence values for predefined outcomes. A prompt contains any of the following elements:
\begin{itemize}
    \item \textbf{Instruction:} a specific task or instruction you want the model to perform
    \item \textbf{Context:} external information or additional context that can steer the model to better responses
    \item \textbf{Input Data:} the input or question that we are interested in finding a response for
    \item \textbf{Output Indicator:} the type or format of the output.
\end{itemize}
You do not need all four elements for a prompt, and the format depends on the task at hand. Although prompting an LLM sounds relatively straightforward, the creation and exact phrasing of a prompt is crucial for the quality of the LLM output. For most tasks, it is necessary to experiment with different prompts and iteratively refine them to yield the best results. This so-called \textit{prompt engineering} step is similar to feature engineering in more traditional ML approaches. Prompt engineering is a relatively new field where only heuristics and simple rules exist that help create effective prompts. Some of these rules are:

\begin{itemize}
    \item \textbf{Start simple:} start with a simple prompt and build on it
    \item \textbf{Call to Action:} start the prompt with an action word like ``Write'', ``Create'', or ``Summarize'' instead of ``Can you''
    \item \textbf{Add Context:} add specific and relevant context to the task you want to perform
    \item \textbf{Add Expectations:} add clear and direct expectations for the content, like how long it should be and what to include
\end{itemize}

Here is an example of a prompt that follows these rules:

\begin{user*}{}
Extract the names of places in the following text. \\
Desired format:\\
Place: <comma\_separated\_list\_of\_places>\\
Input: ``Although these developments are encouraging to researchers, much is still a mystery. ``We often have a black box between the brain and the effect we see in the periphery,'' says Henrique Veiga-Fernandes, a neuroimmunologist at the Champalimaud Centre for the Unknown in Lisbon. ``If we want to use it in the therapeutic context, we actually need to understand the mechanism.'' ''
\end{user*}

\begin{assistant*}{}
Place: Champalimaud Centre for the Unknown, Lisbon
\end{assistant*}

\subsection{Overview of Popular Large Language Models}
\label{sec:LLMs}
Table~\ref{tbl:LLM-overview} contains an overview of the most popular LLMs. The table is structured along their architecture (encoder-only, decoder-only, or encoder-decoder). Please note that each row represents a whole family of models with different instances, which vary in the number of trainable parameters, the size and kind of training data used, support for languages, etc. Platforms like Hugginface\footnote{\url{https://huggingface.co/}} offer searchable repositories to find and download appropriate models for different tasks. 

\begin{table}
  \centering
  \setlength{\tabcolsep}{6pt}
  \setlength{\extrarowheight}{0.1cm}
  \caption{Comparison of Popular Large Language Models}
  \label{tbl:LLM-overview}
  \begin{tabularx}{\textwidth}{l X r r p{1.6cm}@{}}
    \toprule
     \textbf{Name} & \textbf{Description} & \textbf{Year} & \textbf{Parameters} & \textbf{Training Objective} \\
    \midrule
    \multicolumn{5}{@{}l@{}}{\textbf{Encoder-only}}\\
    BERT~\cite{devlin19} & A bidirectionally pre-trained LLM for natural language understanding. & 2018 & 110M--340M & MLM, NSP\\
    RoBERTa~\cite{roberta} & Robustly optimized BERT approach. & 2019 & 125M--355M &MLM\\
    DistilBERT~\cite{distilbert} & A distilled version of BERT with fewer parameters but similar performance. & 2019 & 66M &MLM\textsuperscript{+}\\
    USE~\cite{use} & Universal Sentence Encoder; specialized for embedding entire sentences or paragraphs. & 2018 & 15M--30M &Multi-task*\\
    \multicolumn{5}{@{}l@{}}{\textbf{Decoder-only}} \\ 
    GPT-3~\cite{gpt3} & Generative language model developed by OpenAI. & 2020 & 175B & NWP\\
    %& Claude-2 & An innovative language model with unique features for diverse language tasks. & 2022 & 130B & \\
    LLama-2~\cite{llama2} & Open source generative LLM developed by Meta. & 2023 & 70B & NWP, RLHF\\
    GPT-4~\cite{gpt4} & The successor to GPT-3 with more parameters and more training data & 2023 & N/A & NWP\\
    \multicolumn{5}{@{}l@{}}{\textbf{Encoder-decoder}} \\ 
    T5~\cite{t5} & An LLM specialized on text-to-text translation & 2020 & 60M--11B &MLM\\
    \bottomrule
    \multicolumn{5}{@{}l@{}}{\textbf{MLM}: Masked language modeling}\\
    \multicolumn{5}{@{}l@{}}{\textbf{NSP}: Next sentence prediction}\\
    \multicolumn{5}{@{}l@{}}{\textbf{MLM\textsuperscript{+}}: Masked language modeling + distillation loss + similarity loss}\\
    \multicolumn{5}{@{}l@{}}{\textbf{Multi-task*}: Skip-thought, input-response prediction, and NL inference}\\
    \multicolumn{5}{@{}l@{}}{\textbf{NWP}: Next word prediction}\\
    \multicolumn{5}{@{}l@{}}{\textbf{RLHF}: Reinforcement-learning from human feedback}\\
  \end{tabularx}
\end{table}

\section{How to Use LLMs in RE: A Systematic Guideline}
\label{sec:guideline}

\subsection{Understand and Define the NLP Problem}

Before trying to solve an RE task, a thorough analysis of the task itself is necessary to decide if and how LLMs may be used to solve the tasks. 

\begin{figure}
    \centering
    \includegraphics[width=1\linewidth]{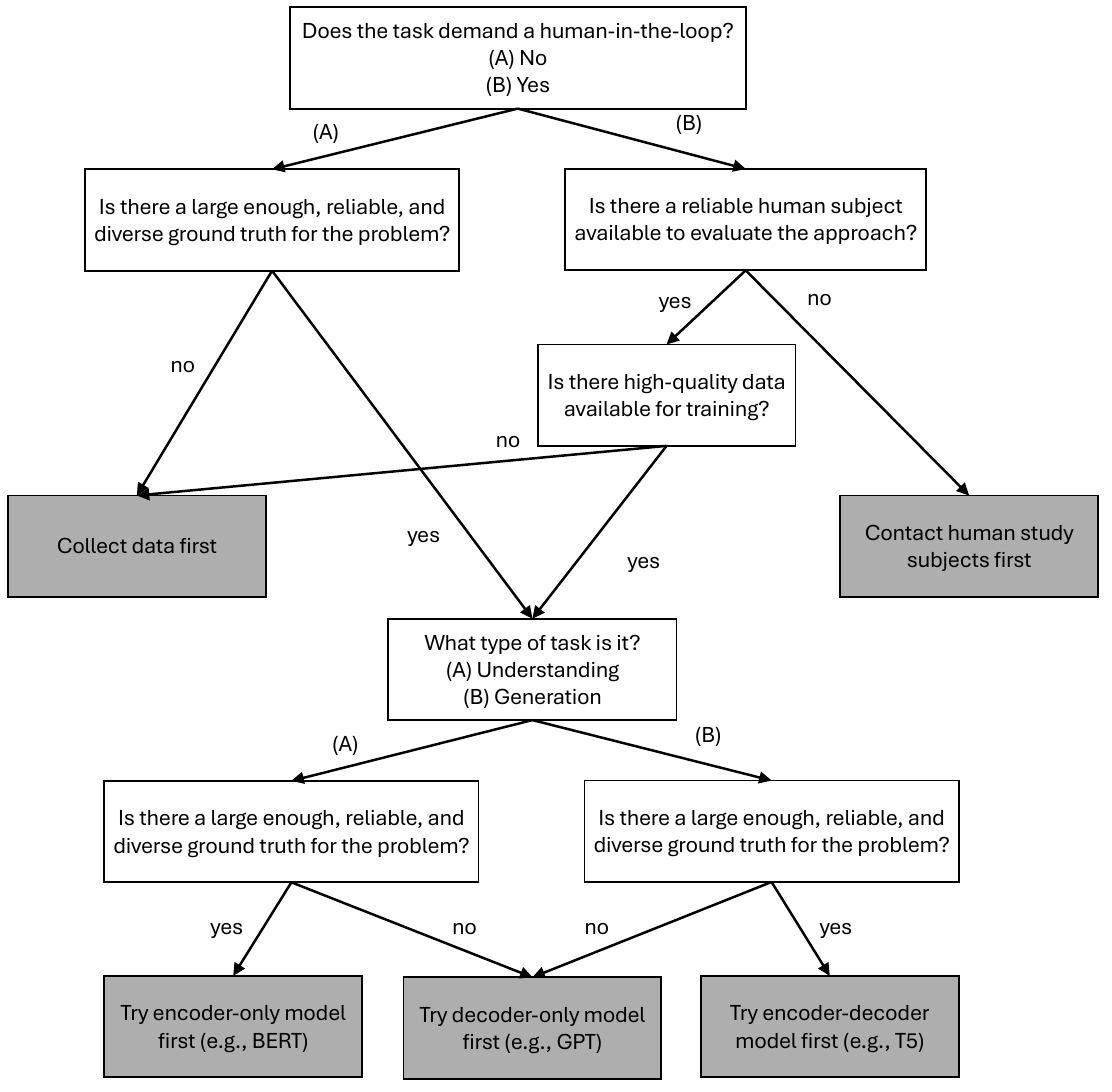}
    \caption{Guiding questions when using LLMs for RE tasks}
    \label{fig:LLM-decisions}
\end{figure}

The upper part of Figure~\ref{fig:LLM-decisions} shows a decision tree that helps decide if a task is ready to be implemented with an LLM. The most important question is related to the envisioned use of the approach either as a fully automated and autonomously acting system or as a supporting system to help humans do their work faster or better. The answer to this question impacts the necessity of high-quality data on the index hand and the necessary type of evaluation and corresponding metrics on the other hand. 

If an approach shall be used to replace a human (i.e., solve a problem autonomously), researchers should make sure that their approach is trained and tested on a large enough, reliable, and diverse ground truth. Diversity is important for ensuring that the approach generalizes to unseen data and that out-of-distribution samples are minimized. A reliable ground truth implies objective and trustworthy labels associated with the data. The approach should be evaluated based on its performance compared to the performance of a human it replaces. If an approach is supposed to support a human in solving a problem (e.g., a recommender system), it is important to develop and evaluate the approach in its context with a human user. The approach should be evaluated based on the (increased) performance of a human using the approach compared to a human without the approach. This means that reliable human subjects are paramount for developing and evaluating such approaches. On the other hand, in such scenarios, a lack of highly reliable and diverse data for training is not that severe since a human can always contextualize and overrule the decisions. Of course, we would expect that better training data will increase the approach's performance and, thus, the value for a human. Once the task is characterized, a researcher may decide how and which LLM to try. The lower part of Figure~\ref{fig:LLM-decisions} shows a decision tree that helps determine what type of LLM may be suitable to solve a particular task. On a high level, RE tasks may be differentiated into \textit{understanding} and \textit{generating} tasks. In both cases, the input is text. 

For \textit{generation} tasks, the output is a text that is generated based on the input. Examples are: 
\begin{itemize}
    \item translation (e.g., requirements to test cases~\cite{Fischbach2023})
    \item summarization (e.g., of stakeholder interviews)
    \item synthesis (e.g., requirements generation~\cite{Koscinski2023})
    \item completion (e.g., of incomplete requirements~\cite{luitel2023improving})
\end{itemize}

For \textit{understanding} tasks, the output is based on an analysis of the input. The input may then be mapped to predefined output classes, clustered concerning each other, or certain input elements may be marked. 
Examples are:
\begin{itemize}
    \item classification (e.g., into types of requirements~\cite{Hey2020})
    \item information retrieval (e.g., glossary term candidates~\cite{Arora2017})
    \item linking (e.g., issues to commits~\cite{lin2021traceability})
    \item clustering (e.g., topic identification~\cite{Gulle2020})
    \item sentiment analysis (e.g., in-app reviews~\cite{Guzman2014})
\end{itemize}

The following three examples illustrate how the above decision trees can be applied.

\begin{tcolorbox}[arc=0mm,width=\columnwidth,
                 top=1mm,left=1mm,  right=1mm, bottom=1mm,
                 boxrule=1pt]
\begin{example} \label{eg1} 
\textbf{Jira ticket classification.} 
An agile development team manages a large number of tickets in Jira. Users or other stakeholders can submit tickets. The development team needs to differentiate different types of tickets quickly. Currently, the product owner reviews every ticket and assigns one of the labels ``bug report'', ``feature request'', or ``other''. The team is looking for an LLM-based solution that automatically assigns the label to newly created Jira tickets, i.e., we follow the path (A) in the decision tree in the upper part of Figure~\ref{fig:LLM-decisions}. Fortunately, we have an existing dataset of more than 300,000 tickets from older projects that have been labeled manually. The development team considers these labels as reliable. Therefore, we can continue with step 2) in the decision tree in the lower part of Figure~\ref{fig:LLM-decisions}. Ticket classification is an understanding task where the LLM needs to understand the input and then map it to predefined classes. Since we have a rather large ground truth at hand, the decision tree suggests trying an encoder-only model (e.g., BERT) first. More information on how this can be done will be given in Section~\ref{sec:encoder-only}.
\end{example}
\end{tcolorbox}

\begin{tcolorbox}[arc=0mm,width=\columnwidth,
                 top=1mm,left=1mm,  right=1mm, bottom=1mm,
                 boxrule=1pt]
\begin{example} \label{eg2} 
\textbf{GDPR compliance of contracts.}
A legal department needs to check whether specific contracts are GDPR compliant. This is a manual and laborious task. The department is looking for automation that helps the team members make this decision, i.e., we follow the path (B) in the decision tree in the upper part of Figure~\ref{fig:LLM-decisions}. Fortunately, the legal department is highly interested in the solution and is willing to evaluate approaches. The legal department has a set of contracts, including their compliance verdict, but this set is not very large. Additionally, contracts are highly confidential and should not be used outside the organization. GDPR compliance checking is an understanding task where the LLM needs to understand the input and then decide on compliance. Since we only have a rather limited ground truth at hand, the decision tree suggests trying a decoder-only model (e.g., GPT) first. More information on how this can be done will be given in Section~\ref{sec:decoder-only}.
\end{example}
\end{tcolorbox}

\begin{tcolorbox}[arc=0mm,width=\columnwidth,
                 top=1mm,left=1mm,  right=1mm, bottom=1mm,
                 boxrule=1pt]
\begin{example} \label{eg3} 
\textbf{Test case generation from use cases.}
A testing team creates test case descriptions based on use case descriptions. This is a manual and laborious task where even experts sometimes forget to create test cases for specific flows of a use case. The team is looking for automation that supports the team members in creating test cases from use case descriptions, i.e., we follow the path (B) in the decision tree in the upper part of Figure~\ref{fig:LLM-decisions}. The testing team is highly interested in the solution and is willing to evaluate approaches. From former projects, the team has access to 500 use cases and 1,500 test cases manually crafted for the use cases. 
The development team considers this dataset to be reliable since the artifacts have undergone an extensive review process. Therefore, we can continue with step 2) in the decision tree in the lower part of Figure~\ref{fig:LLM-decisions}. Test case generation is a generation task where the LLM processes the input and then generates corresponding test cases. Since we have a rather large and reliable ground truth at hand, the decision tree suggests trying an encoder-decoder model (e.g., T5) first. More information on how this can be done will be given in Section~\ref{sec:encoder-decoder}.
\end{example}
\end{tcolorbox}

\subsection{LLM Architectures and their Usage}
\begin{figure}
    \centering
    \includegraphics[width=1\linewidth]{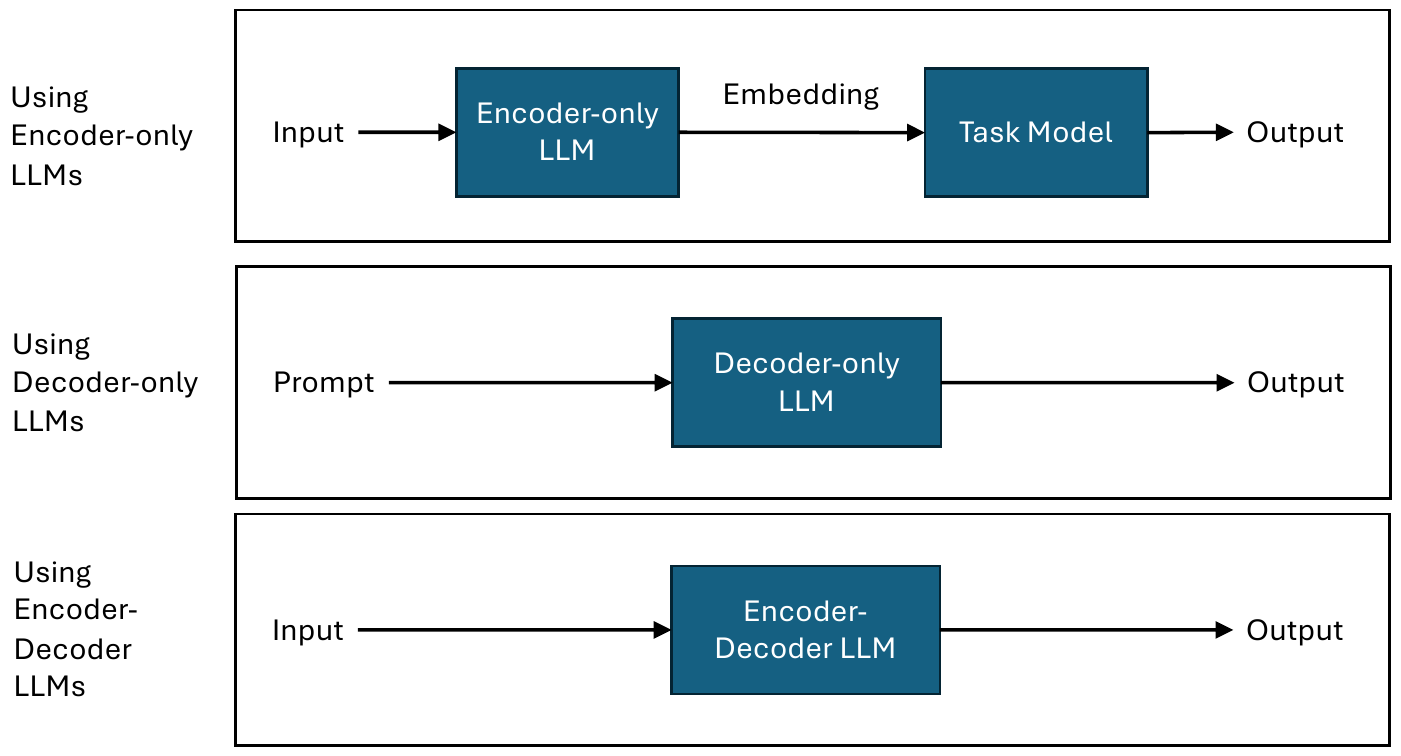}
    \caption{Simplified overview of using different LLM architectures to solve tasks}
    \label{fig:usages}
\end{figure}

As described in Section~\ref{sec:fundamentals}, an LLM's architecture defines how input is processed and what type of output the LLM generates. These differences impact the way how an LLM must be used to solve a certain task. Figure~\ref{fig:usages}) provides a schematic overview of how input and output change depending on which architecture is used to solve a task.
Encoder-only LLMs like BERT (see Section~\ref{sec:LLMs}) are primarily designed to provide high-quality language embeddings, which can be used as input for the intended task. 
Decoder-only LLMs like GPT provide another option to solve NLP4RE tasks. The generative capabilities can prompt these models with the original tasks and use the outputs as results. 
Encoder-decoder LLMs like T5 are specialized for text-to-text translation. 
In the following, we will explain these usage modes in detail.

\subsection{Using Encoder-only LLMs}
\label{sec:encoder-only}
\begin{figure}
    \centering
    \includegraphics[width=0.8\linewidth]{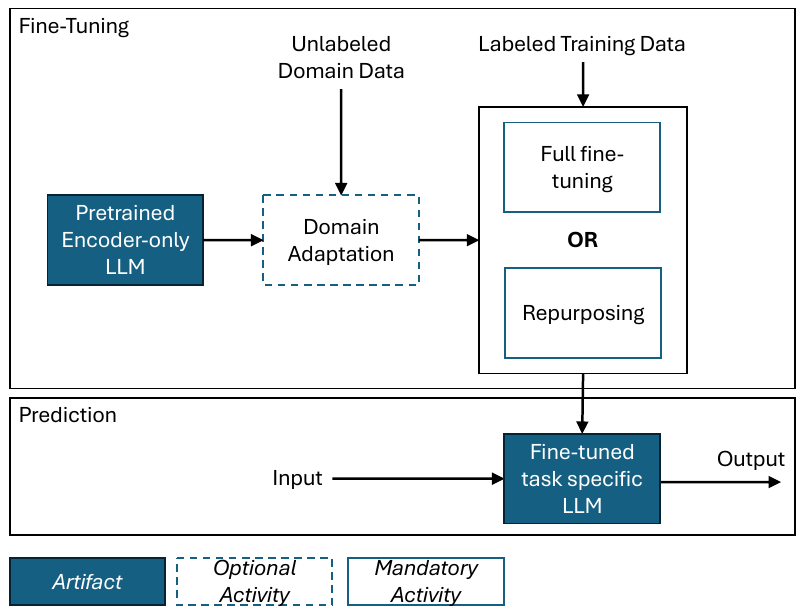}
    \caption{Schematic pipeline when using encoder-only LLMs}
    \label{fig:encoder-only}
\end{figure}

Figure~\ref{fig:encoder-only} shows potential pipelines using the LLMs as language embedders. In this usage mode, a specific RE task is solved by leveraging the semantically rich embeddings generated by an LLM. In contrast to simpler embedding techniques like \textit{tf-idf} or \textit{bag-of-words}, LLM embeddings better capture the meaning of a word in its context (see Section~\ref{sec:embeddings}). The embeddings are then used in specific task models to solve the actual RE task (e.g., classification, clustering, transformation). This step is called \textit{repurposing} because it changes the purpose of the generic LLM to solve the desired task. 

\textbf{Repurposing (Task Model).}
Adjusting an LLM to solve a specific task is also called \textit{repurposing} the model. Repurposing is usually done by adding additional output layers to the pre-trained LLM. That means the task model is not separate but an addition to the pre-trained LLM. For example, you may add a fully connected dense layer with one output neuron per prediction class to solve a classification task. The model is then trained based on a set of labeled training data. In the training process, only the parameters of the additional output layers are adjusted. Therefore, repurposing can be done with relatively little data.

An example of this pipeline is the work by Hey~et~al.~\cite{Hey2020}. In their work, they repurpose a pre-trained BERT model based on the PROMISE NFR dataset~\cite{PROMISE_NFR}, which contains requirements and their associated classes (functional, usability, performance, security, operational). They repurpose the model by adding a single layer of linear neurons in a feedforward neural
network. The outputs are directly computed from the sum of the weighted inputs (plus some bias). They use the softmax function to get a probability distribution for the different labels. The quality of the provided output depends on the embeddings for the given task. In the following, we describe techniques that can be used to further refine the generated embeddings for the given task.

\textbf{Domain Adaptation.}
The largest and most powerful LLMs are general-purpose LLMs, i.e., they are trained on all kinds of publicly available data. While this ensures generalizability, a pre-trained LLM may not be able to capture the specifics of a certain domain. This is especially relevant if the tasks that shall be solved contain a lot of technical or even company-specific jargon. 
Unsupervised fine-tuning can adapt a pre-trained LLM to a specific domain. The general idea is to continue training the pre-trained LLM on a corpus of unlabeled but domain-specific data. 
For example, Chang~et~al.~\cite{Chang2023} describe an approach for automated requirements linking that starts from a BERT-based model pre-trained in one particular domain and is then adapted to another.  
Domain-specific data can be acquired in several ways. For example, Ezzini~et~al.\ presented WikiDoMiner, which builds a domain-specific corpus from Wikipedia based on keywords found in some input documents~\cite{Ezzini22}. Unsupervised fine-tuning can also be used to incorporate the knowledge contained in sensitive company documents. 

\textbf{Supervised (Full) Fine-Tuning.}
Supervised Full Fine-Tuning (SFFT) combines repurposing and domain adaptation. In SFFT, a labeled dataset is used to repurpose the LLM by adding output layers; however, the parameters of the attention layers (i.e., the transformer model) can also be updated during fine-tuning. 
This operation can be computationally expensive and complicated, depending on the size of your model. In some cases, you can keep parts of the transformer model frozen to reduce fine-tuning costs.

\subsection{Using Decoder-only LLMs with Prompts}
\label{sec:decoder-only}
\begin{figure}
    \centering
    \includegraphics[width=0.88\linewidth]{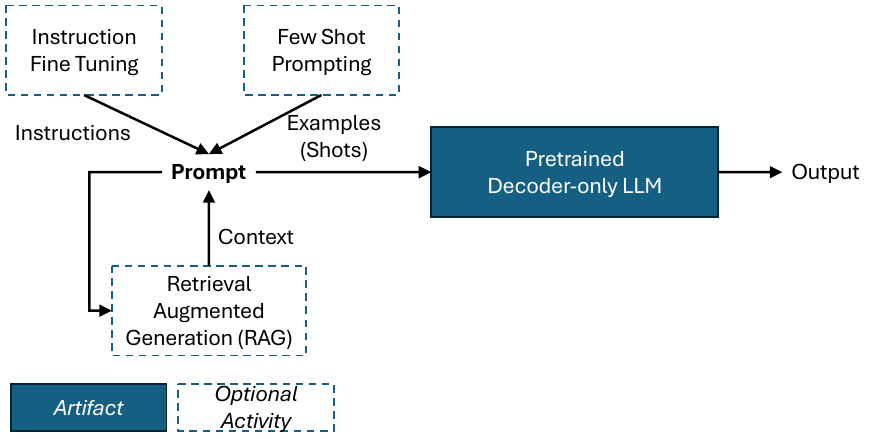}
    \caption{Schematic pipeline when using decoder-only LLMs}
    \label{fig:decoder-only}
\end{figure}

Figure~\ref{fig:decoder-only} shows potential pipelines when using the generative capabilities of LLMs. In this usage mode, a specific RE task is written as a prompt (see Section\ref{sec:prompting}), which is input to a generative LLM. The output of the model directly provides the answer to the specific task. 

Here is an example of how a trace link recovery task can be formulated as a prompt~\cite{Rodriguez23,Traceability_Chapter}:

\begin{user*}{}
Below are artifacts from the same software system. Is there a traceability link between (1) and (2)?
\vspace{1ex}

(1) The DPU-TMALI shall utilize SCM\_DCI\_SR, along with ERRNO provided by DPU-DCI to decode errors and place them on an error queue for DPU-CCM.
\vspace{1ex}

(2) Error Collection and Reporting The ccmErrEnq() function tracks the last error reported and its frequency of occurrence. Once an error code has been reported it becomes the previously reported error code maintained by ccmErrEnq(). A repetition count is then incremented for each subsequent, consecutively reported, identical instance of this previously reported error. If this error code is reported more than once in one high-rate housekeeping reporting period, then a special error, S\_ccm\_ERR\_REPEAT is enqueued with the repetition count for the error encoded in the least significant byte. This mechanism effectively reduces the potential for housekeeping telemetry to become flooded with a single repeated error.
\end{user*}

\begin{assistant*}{}
Yes, there is a traceability link between (1) and (2). 
(1) mentions that the DPU-TMALI places errors on an error queue for the DPU-CCM.
\vspace{1ex}

(2) describes the error collection and reporting mechanism of the DPU-CCM, specifically referring to enqueuing errors.
\vspace{1ex}

So (2) provides an implementation detail for the error queuing functionality briefly mentioned in (1). Therefore, there is a traceability link between these two artifacts in the software system.
\end{assistant*}

The quality of the provided output depends on the specific prompt. Prompts can be refined by refining any component (i.e., instructions, context, input question). For example, we could change the prompt in the above example to:

\begin{user*}{}
Is (1) related to (2)? Answer ``yes'' or ``no''.
%\vspace{1ex}

(1) \ldots
%\vspace{1ex}

(2) \ldots
\end{user*}

This input question makes the output less wordy and easier to process. Reformulating the prompt can help generate answers that fit an expected format. For increasing the semantic quality of the output, providing context and instructions is a more effective strategy. In the following, we describe three approaches that can be used to refine the prompt and increase the model's performance.

\textbf{Few-Shot Prompting.}
In few-shot prompting, a prompt is enriched with some examples of the problem to be solved. The examples may stem from existing (training) data or can be artificial. In the example above, a few-shot prompting strategy may look like this:

\begin{user*}{}
Requirements:

(a) \ldots 

(b) \ldots

(c) \ldots

(d) \ldots

\vspace{1ex}

(a) is related only to (b).

(b) is related to (a) and (c).

(c) is related to (b) and (d).

\vspace{1ex}

Is (1) related to (2)? Answer ``yes'' or ``no''.

(1) \ldots

(2) \ldots
\end{user*}

The examples, also called \textit{shots}, help the LLM to identify the context of the task better and learn the meaning of specific concepts, such as traceability. The closer the examples are to the problem to be solved, the more the LLM can profit from them. 

\textbf{Instruction Fine-Tuning.}
Besides providing examples, prompts can also incorporate \textit{instructions}, which help the LLM understand the task and how to solve it (see Section~\ref{sec:prompting}). Instructions may contain certain keywords such as \textit{classify}, \textit{summarize}, or \textit{translate}, which define the type of task. The instruction may also contain information about the output format and style

Here is an example of a prompt for trace link recovery as suggested by Rodriguez~et~al.~\cite{Rodriguez23}.

\begin{user*}{}
\# Instructions: \\
Rank all related artifacts from most to least related to the source. Provide the ranked artifacts as a comma-delimited list of artifact IDs where the first element relates to the source the most and the last element does so the least.
\vspace{1ex}

\# Source: [SOURCE ARTIFACT] 
\vspace{1ex}

\# Artifacts:\\
\ldots
\end{user*}

\begin{figure}
    \centering
    \includegraphics[width=1\linewidth]{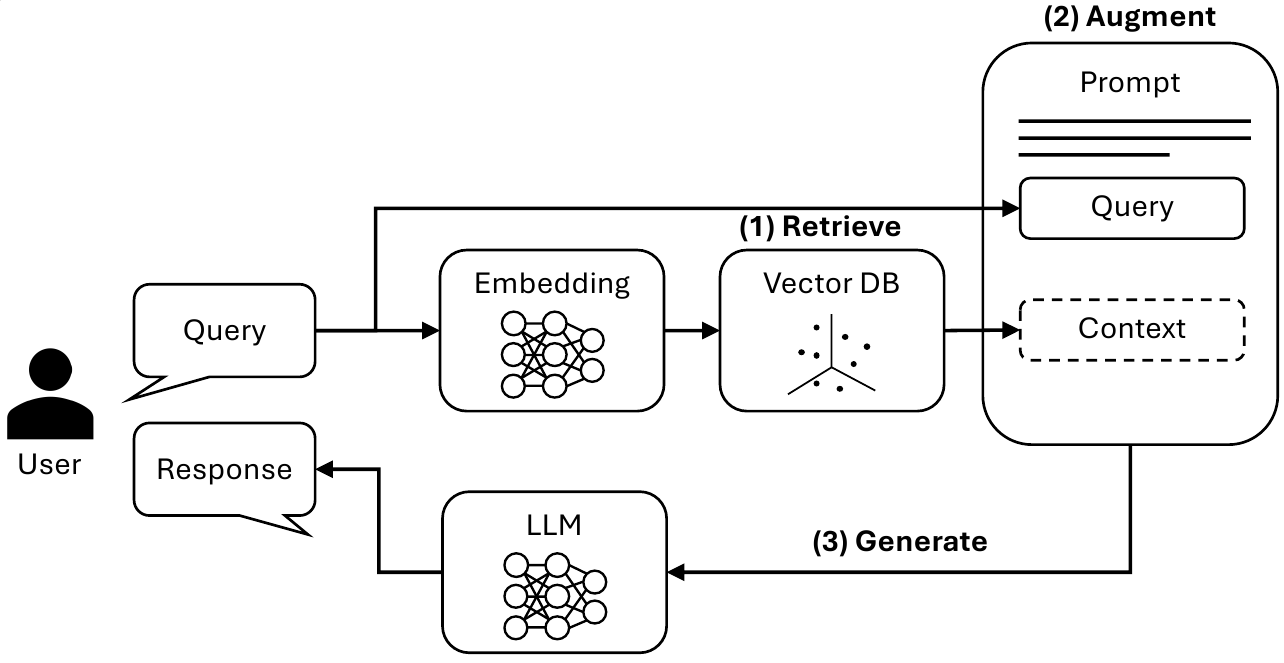}
    \caption{Retrieval Augmented Generation (RAG)}
    \label{fig:rag}
\end{figure}

\textbf{Retrieval Augmented Generation (RAG).}
Instructions and examples (shots) are a way to provide context to the LLM. Providing context is the most effective way to optimize the results when working with generative LLMs. However, the capabilities of LLMs to process context is limited in size. For example, GPT-4 and Llama 2 accept a maximum of 8,000 and 4,000 tokens as context. A special version of GPT-4 accepts up to \textasciitilde 32,000 tokens (\textasciitilde 3,000 words) as context. Although LLMs will likely increase this maximum in the future, it may still be impossible to provide large sets of artifacts as context (e.g., an entire requirements specification). Retrieval Augmented Generation (RAG) is a technique that inserts a sample from a large context corpus that is most likely to fit a certain task. Figure~\ref{fig:rag} provides an overview of the approach. 

RAG takes an input (a query) and retrieves a set of relevant/supporting documents by embedding the query into a vector space that contains embeddings of documents from a given source (e.g., Wikipedia or company-specific documents). The most relevant documents from the vector DB are concatenated as context with the original query to form a prompt fed to the decoder-only LLM, producing the final output. This makes RAG adaptive for situations where facts could evolve.  
One useful design pattern is to create a vector database that stores embeddings of company documents. When the user enters a prompt, the vector DB retrieves relevant documents and sends them as context to the model. RAG allows language models to bypass retraining, enabling access to the latest information for generating reliable outputs via retrieval-based generation.

\subsection{Using Encoder-Decoder LLMs}
\label{sec:encoder-decoder}
\begin{figure}
    \centering
    \includegraphics[width=0.7\linewidth]{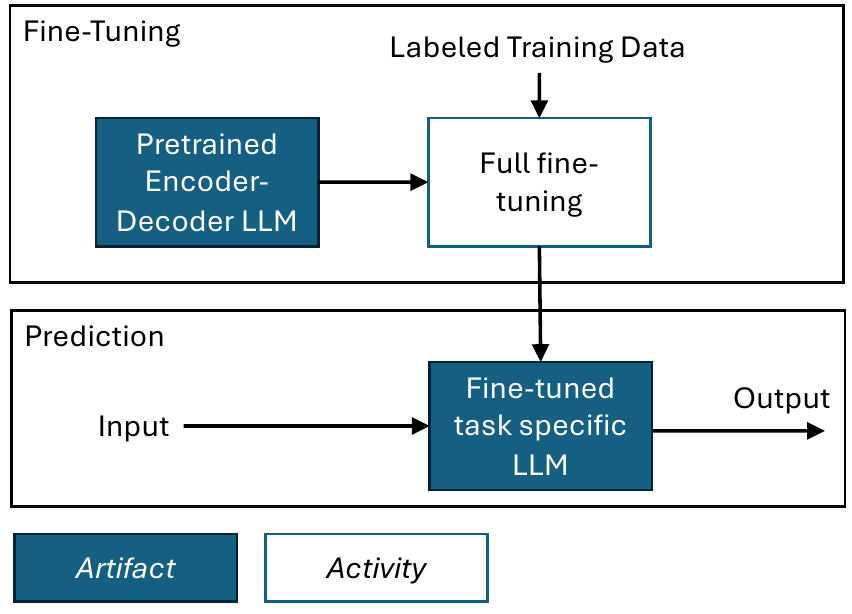}
    \caption{Schematic pipeline when using encoder-decoder LLMs}
    \label{fig:encoder-decoder}
\end{figure}

Encoder-decoder LLMs are designed for \textit{sequence-to-sequence} tasks. They translate sequences of textual input to sequences of textual output. In contrast to decoder-only LLMs, the input sequence does not represent a prompt but a genuine input that needs to be translated to a genuine output of unknown lengths. Therefore, encoder-decoder LLMs are closer to encoder-only LLMs than to decoder-only LLMs when it comes to fine-tuning. As shown in Figure~\ref{fig:encoder-decoder}, when working with pre-trained encoder-decoder LLMs, fine-tuning is usually done by \textit{fully fine-tuning} the pre-trained LLM with a training data set that contains input sequences and corresponding target sequences. Instead of working with pre-trained encoder-decoder LLMs such as T5, you can also build an encoder-decoder LLM by combining a pre-trained encoder-only LLM (e.g., BERT) with a pre-trained decoder-only LLM (e.g., GPT). 

\section{Summary and Conclusion}
\label{sec:conclusion}

In this chapter, we have introduced the fundamentals of large language models (LLMs) and shown different ways to use pre-trained LLMs for solving RE tasks. For most RE tasks, pre-trained LLMs are the cornerstone for high-performing automation because they encode a high level of general language understanding, which is the basis for most NLP4RE tasks. In this way, pre-trained LLMs have revolutionized the possibilities for automation in RE where high-quality data is usually too scarce to train language models from scratch. 

The main challenge when working with LLMs is the fine-tuning step, where the pre-trained LLM is adapted to solve a specific task it has not been trained on. We support this challenge with a decision tree (see Figure~\ref{fig:LLM-decisions}) that helps identify which type of LLM is most amendable for the problem to be solved. We show the different fine-tuning possibilities for each type of LLM.
Fine-tuning, however, also bears specific challenges that should be considered and evaluated. These include:
\begin{itemize}
    \item \textbf{Overfitting}: Fine-tuning may lead to models that become too specific to the training data, leading to suboptimal generalization on unseen data. Regularization techniques such as dropout and weight decay can be applied during fine-tuning to prevent overfitting. Additionally, carefully curating the training data and utilizing techniques like cross-validation can help maintain a balance between model complexity and generalization. You can also consider stopping early, ensemble learning, regularly auditing performance, and monitoring model complexity.
    \item \textbf{Bias amplification}: Explicit or implicit biases in training data get amplified during training. Since fine-tuning is a continuation of the training process, it amplifies potential biases even further.
    \item \textbf{Hyperparameter tuning}: The process of fine-tuning involves various hyperparameters that govern the optimization process, such as learning rates, batch sizes, and regularization strengths. Hyperparameters heavily influence the performance of a fine-tuned LLM. Selecting inappropriate hyperparameters can lead to slow convergence, poor generalization, or even unstable training.
\end{itemize}

In some cases, LLM fine-tuning is not possible or not useful: 

\begin{itemize}
    \item Some models are only available through application programming interfaces (API) with no or limited fine-tuning services.
    \item You might not have enough data to fine-tune the model for the downstream task or your application's domain.
    \item The data in the application might change frequently. Fine-tuning the model frequently might not be possible or might be detrimental. For example, the data in news-related applications changes every day.
    \item The application might be dynamic and context-sensitive. For example, if you create a chatbot that customizes its output for each user, you cannot fine-tune the model on user data.
\end{itemize}

The fine-tuning possibilities mentioned in this chapter are not complete. The fact that most LLMs are provided as more or less ``open'' models makes numerous fine-tuning methods possible. For example, it is also possible to fine-tune the weights of decoder-only models instead of just fine-tuning them by tuning the prompts. Similarly, encoder-decoder models may also undergo an unsupervised fine-tuning step with unlabeled domain documents. It is also important to understand that one RE task cannot necessarily only be solved with one LLM architecture. For example, decoder-only LLMs such as GPT have also shown remarkable results for text translation tasks, which seem more amendable to encoder-decoder models. Similarly, encoder-decoder models have also been used for text understanding tasks such as review classification. Therefore, working with LLMs is inherently experimental. Researchers may use the content of this chapter as a guideline but should always experiment with alternatives. It is also important to keep an eye on the quickly evolving landscape of LLMs. As shown in Table~\ref{tbl:LLM-overview}, new LLMs are constantly being published. Platforms for sharing and publishing pre-trained models accelerate this development. Huggingface, one of the most popular repositories for pre-trained models, currently\footnote{in December 2023} hosts over 40,000 pre-trained models just for text classification. Many of the provided models are specialized in certain tasks, certain languages, or certain domains. Whereas most authors in RE use the basic versions of a particular LLM (e.g., \texttt{bert-base-uncased}\footnote{\url{https://huggingface.co/bert-base-uncased}} from the BERT family or \texttt{t5-base}\footnote{\url{https://huggingface.co/google-t5}} from the T5 family), it may be interesting to consider more specific pre-trained alternatives available on pre-trained model repositories.

%

 \bibliographystyle{splncs04}
 \bibliography{references}

\begin{thebibliography}{10}
\providecommand{\url}[1]{\texttt{#1}}
\providecommand{\urlprefix}{URL }
\providecommand{\doi}[1]{https://doi.org/#1}

\bibitem{Arora2017}
Arora, C., Sabetzadeh, M., Briand, L., Zimmer, F.: Automated extraction and
  clustering of requirements glossary terms. IEEE Transactions on Software
  Engineering  \textbf{43}(10),  918--945 (2017).
  \doi{10.1109/tse.2016.2635134}

\bibitem{Bengio13}
Bengio, Y., Courville, A., Vincent, P.: Representation learning: A review and
  new perspectives. IEEE Transactions on Pattern Analysis and Machine
  Intelligence  \textbf{35}(8),  1798--1828 (2013). \doi{10.1109/TPAMI.2013.50}

\bibitem{blitzer06}
Blitzer, J., McDonald, R., Pereira, F.: Domain adaptation with structural
  correspondence learning. In: Conference on Empirical Methods in Natural
  Language Processing. pp. 120--128. Association for Computational Linguistics,
  Sydney, Australia (2006)

\bibitem{bojanowski16}
Bojanowski, P., Grave, E., Joulin, A., Mikolov, T.: Enriching word vectors with
  subword information. CoRR  \textbf{abs/1607.04606} (2016)

\bibitem{gpt3}
Brown, T., Mann, B., Ryder, N., Subbiah, M., Kaplan, J.D., Dhariwal, P.,
  Neelakantan, A., Shyam, P., Sastry, G., Askell, A., et~al.: Language models
  are few-shot learners. Advances in neural information processing systems
  \textbf{33},  1877--1901 (2020)

\bibitem{use}
Cer, D., Yang, Y., Kong, S.y., Hua, N., Limtiaco, N., John, R.S., Constant, N.,
  Guajardo-Cespedes, M., Yuan, S., Tar, C., Sung, Y.H., Strope, B., Kurzweil,
  R.: Universal sentence encoder (2018). \doi{10.48550/ARXIV.1803.11175}

\bibitem{Chang2023}
Chang, Z., Li, M., Wang, Q., Li, S., Wang, J.: Cross-domain requirements
  linking via adversarial-based domain adaptation. In: IEEE/ACM 45th
  International Conference on Software Engineering (ICSE). IEEE (2023).
  \doi{10.1109/icse48619.2023.00138}

\bibitem{PROMISE_NFR}
Cleland-Huang, J., Mazrouee, S., Liguo, H., Port, D.: {NFR} (2007).
  \doi{10.5281/ZENODO.268542}

\bibitem{devlin19}
Devlin, J., Chang, M.W., Lee, K., Toutanova, K.: {BERT}: Pre-training of deep
  bidirectional transformers for language understanding (2018).
  \doi{10.48550/ARXIV.1810.04805}

\bibitem{Ezzini22}
Ezzini, S., Abualhaija, S., Sabetzadeh, M.: {WikiDoMiner}: Wikipedia
  domain-specific miner. In: 30th ACM Joint European Software Engineering
  Conference and Symposium on the Foundations of Software Engineering
  (ESEC/FSE). pp. 1706--1710. Association for Computing Machinery, New York,
  NY, USA (2022). \doi{10.1145/3540250.3558916}

\bibitem{Firth57}
Firth, J.R.: A synopsis of linguistic theory. Oxford: Philological Society
  (1957), reprinted in F. Palmer (ed.)(1968). Studies in Linguistic Analysis
  1930-1955. Selected Papers of J.R. Firth., Harlow: Longman

\bibitem{Fischbach2023}
Fischbach, J., Frattini, J., Vogelsang, A., Mendez, D., Unterkalmsteiner, M.,
  Wehrle, A., Henao, P.R., Yousefi, P., Juricic, T., Radduenz, J., Wiecher, C.:
  Automatic creation of acceptance tests by extracting conditionals from
  requirements: {NLP} approach and case study. Journal of Systems and Software
  \textbf{197} (2023). \doi{10.1016/j.jss.2022.111549}

\bibitem{Gulle2020}
G{\"u}lle, K.J., Ford, N., Ebel, P., Brokhausen, F., Vogelsang, A.: Topic
  modeling on user stories using word mover’s distance. In: IEEE Seventh
  International Workshop on Artificial Intelligence for Requirements
  Engineering (AIRE). IEEE (2020). \doi{10.1109/aire51212.2020.00015}

\bibitem{Traceability_Chapter}
Guo, J.L.C., Stegh{\"o}fer, J.P., Vogelsang, A., Cleland-Huang, J.: Natural
  language processing for requirements traceability. In: Ferrari, A.,
  Deshpande, G. (eds.) Handbook of Natural Language Processing for Requirements
  Engineering. Springer International Publishing, Cham (2024)

\bibitem{Guzman2014}
Guzman, E., Maalej, W.: How do users like this feature? a fine grained
  sentiment analysis of app reviews. In: IEEE 22nd International Requirements
  Engineering Conference (RE). IEEE (2014). \doi{10.1109/re.2014.6912257}

\bibitem{Hey2020}
Hey, T., Keim, J., Koziolek, A., Tichy, W.F.: Norbert: Transfer learning for
  requirements classification. In: IEEE 28th International Requirements
  Engineering Conference (RE). IEEE (2020). \doi{10.1109/re48521.2020.00028}

\bibitem{Jurafsky2009}
Jurafsky, D., Martin, J.H.: Speech and language processing: an introduction to
  natural language processing, computational linguistics, and speech
  recognition. Pearson Prentice Hall, Upper Saddle River, N.J. (2009)

\bibitem{Koscinski2023}
Koscinski, V., Hashemi, S., Mirakhorli, M.: On-demand security requirements
  synthesis with relational generative adversarial networks. In: IEEE/ACM 45th
  International Conference on Software Engineering (ICSE). IEEE (2023).
  \doi{10.1109/icse48619.2023.00139}

\bibitem{lin2021traceability}
Lin, J., Liu, Y., Zeng, Q., Jiang, M., Cleland-Huang, J.: Traceability
  transformed: Generating more accurate links with pre-trained {{BERT}} models.
  In: {IEEE}/{ACM} 43rd International Conference on Software Engineering
  ({ICSE}). {IEEE} (2021). \doi{10.1109/icse43902.2021.00040}

\bibitem{roberta}
Liu, Y., Ott, M., Goyal, N., Du, J., Joshi, M., Chen, D., Levy, O., Lewis, M.,
  Zettlemoyer, L., Stoyanov, V.: {RoBERTa}: A robustly optimized {BERT}
  pretraining approach (2019). \doi{10.48550/ARXIV.1907.11692}

\bibitem{luitel2023improving}
Luitel, D., Hassani, S., Sabetzadeh, M.: Improving requirements completeness:
  Automated assistance through large language models. Requirements Engineering
  Journal (REJ)  (2023)

\bibitem{Mikolov13}
Mikolov, T., Sutskever, I., Chen, K., Corrado, G.S., Dean, J.: Distributed
  representations of words and phrases and their compositionality. In: Burges,
  C.J.C., Bottou, L., Welling, M., Ghahramani, Z., Weinberger, K.Q. (eds.)
  Advances in Neural Information Processing Systems. vol.~26. Curran
  Associates, Inc. (2013)

\bibitem{gpt4}
OpenAI:~Achiam, J., Adler, S., Agarwal, S., Ahmad, L., Akkaya, I., Aleman,
  F.L., Almeida, D., \emph{et al.}: {GPT}-4 technical report (2023).
  \doi{10.48550/ARXIV.2303.08774}

\bibitem{Osgood57}
Osgood, C.E., Suci, G.J., Tannenbaum, P.H.: The Measurement of Meaning.
  University of Illinois Press (1957)

\bibitem{penningto14}
Pennington, J., Socher, R., Manning, C.: {G}lo{V}e: Global vectors for word
  representation. In: Conference on Empirical Methods in Natural Language
  Processing ({EMNLP}). pp. 1532--1543. Association for Computational
  Linguistics, Doha, Qatar (2014). \doi{10.3115/v1/D14-1162}

\bibitem{t5}
Raffel, C., Shazeer, N., Roberts, A., Lee, K., Narang, S., Matena, M., Zhou,
  Y., Li, W., Liu, P.J.: Exploring the limits of transfer learning with a
  unified text-to-text transformer (2019). \doi{10.48550/ARXIV.1910.10683}

\bibitem{Rodriguez23}
Rodriguez, A.D., Dearstyne, K.R., Cleland-Huang, J.: Prompts matter: Insights
  and strategies for prompt engineering in automated software traceability. In:
  IEEE 31st International Requirements Engineering Conference Workshops (REW).
  pp. 455--464 (2023). \doi{10.1109/REW57809.2023.00087}

\bibitem{distilbert}
Sanh, V., Debut, L., Chaumond, J., Wolf, T.: {DistilBERT}, a distilled version
  of {BERT}: smaller, faster, cheaper and lighter (2019).
  \doi{10.48550/ARXIV.1910.01108}

\bibitem{wordpiece}
Schuster, M., Nakajima, K.: Japanese and korean voice search. In: International
  Conference on Acoustics, Speech and Signal Processing. pp. 5149--5152 (2012)

\bibitem{llama2}
Touvron, H., Martin, L., Stone, K., Albert, P., Almahairi, A., Babaei, Y.,
  Bashlykov, N., Batra, S., Bhargava, P., Bhosale, S., Bikel, D., Blecher, L.,
  Ferrer, C.C., Chen, M., Cucurull, G., Esiobu, D., Fernandes, J., Fu, J., Fu,
  W., Fuller, B., Gao, C., Goswami, V., Goyal, N., Hartshorn, A., Hosseini, S.,
  Hou, R., Inan, H., Kardas, M., Kerkez, V., Khabsa, M., Kloumann, I., Korenev,
  A., Koura, P.S., Lachaux, M.A., Lavril, T., Lee, J., Liskovich, D., Lu, Y.,
  Mao, Y., Martinet, X., Mihaylov, T., Mishra, P., Molybog, I., Nie, Y.,
  Poulton, A., Reizenstein, J., Rungta, R., Saladi, K., Schelten, A., Silva,
  R., Smith, E.M., Subramanian, R., Tan, X.E., Tang, B., Taylor, R., Williams,
  A., Kuan, J.X., Xu, P., Yan, Z., Zarov, I., Zhang, Y., Fan, A., Kambadur, M.,
  Narang, S., Rodriguez, A., Stojnic, R., Edunov, S., Scialom, T.: Llama 2:
  Open foundation and fine-tuned chat models (2023).
  \doi{10.48550/ARXIV.2307.09288}

\bibitem{NIPS2017}
Vaswani, A., Shazeer, N., Parmar, N., Uszkoreit, J., Jones, L., Gomez, A.N.,
  Kaiser, {\L}., Polosukhin, I.: Attention is all you need. Advances in neural
  information processing systems  \textbf{30} (2017)

\bibitem{Winkler2016}
Winkler, J., Vogelsang, A.: Automatic classification of requirements based on
  convolutional neural networks. In: IEEE 24th International Requirements
  Engineering Conference Workshops (REW). IEEE (2016).
  \doi{10.1109/rew.2016.021}

\end{thebibliography}

\end{document}